Goldsmiths University of London

Computing Department

MS.c User Experience Engineering, 2023

# Improving Digital Mentorship: Insights and Recommendations from the Re:Coded Community Platform Case Study

Author:

Huda Maytham Najm Alabbas

Academic Supervisor:

Dr Yoram Chisik

## Abstract:

With the rapid growth of technology, emerging IT professionals increasingly require mentorship to secure positions in the field. Recognising this need, Re:Coded has enhanced the skill set of their tech bootcamp graduates by introducing the community platform - “Re:Coded’s Mentorship Platform”.

To improve the user experience of the volunteer mentors at Re:Coded, this thesis investigates optimising mentors' interactions with digital mentorship platforms and suggestions for enhancing these interactions.

Multiple third-party collaborators have powered the mentorship platform at Re:Coded. This thesis examines the platform powered by StellarUp as a case study. The insights obtained may inform the UX design of any subsequent mentorship platforms considered by Re:Coded.

This thesis adopted a user-centric approach to solving a UX question. The study identified challenges in the mentors' user journey and their needs by engaging with users via interviews, usability tests, and eye-tracking methods.

Three principal issues emerged: platform navigation, the onboarding process, and the seamless integration of external tools. Solutions were derived from desk research for onboarding, card sorting techniques for navigation, and competitive analyses for tool integration.

Throughout the research, feedback from 23 participants was gathered, ensuring a holistic understanding and actionable recommendations for developing a user-friendly and efficient mentorship platform.

## Acknowledgements:

I am genuinely grateful to my respectful supervisor, Dr Yoram Chisik, for his endless support throughout this journey. This project was not the smoothest, but his guidance and belief made it more accessible.

My father, Maytham Alabbas, for his support and kind words throughout this year, and my brother, Mohammed Alabbas, for his encouragement, which drove me to undertake the most critical life decisions, including this one. I would never be here today without him.

My dearest friends, especially my best friend, Areeg Fahad, have been incredibly supportive at every stage of my professional life, always there to provide encouragement and support when needed.

I would like to express my gratitude to Ben Miller and Elia Maniscalco from Re:Coded for their support throughout my entire academic year, particularly their invaluable assistance during this project.

The honourable Chevening Scholarship team, who saw a potential in me among thousands of candidates and awarded me the incredible opportunity to realise my dreams.

I am grateful to my friends and colleagues at Goldsmiths for the fantastic atmosphere of sharing knowledge and perspectives.

Lastly, I dedicate this achievement to my beloved mother, Baydaa Hassan. She instilled in me that education and curiosity are essential in life. Her pure soul and ambition have always inspired me to empower myself and fight for my dreams. This one is for you, Mom.

## Table of Contents:

# Chapter 1: Introduction

The fast growth of technology in recent years has resulted in a remarkable rise in the need for experienced tech experts (Kwan, 2002). Recognising this rising need, organisations such as Re:Coded have adopted a proactive approach, stepping in to fill the need. They have positioned themselves as more than educational institutions, developing the next generation of digital specialists and leaders. In a bold move in 2022, Re:Coded embarked on an ambitious expansion, training over 100 students in each cohort. This change demonstrates their constant commitment to the IT sector's progress. It reflects their more significant aim of remaining at the forefront of digital transformation, fostering talent, and ensuring the information technology community flourishes.

In 2023, the organisation adopted an innovative approach to support the scaled numbers of graduates. Re:Coded has embarked on launching a new mentorship program and a new platform called 'The Community Platform' to manage this program. Two critical goals are central to this attempt: first, to ensure that mentors are fully prepared with the necessary abilities to guide students effectively, and second, to build a satisfying user experience for mentors using the platform, developing their passion, especially since these mentors involved in this job as volunteers. By fulfilling these objectives, Re:Coded hopes to increase mentor loyalty and motivate them to become passionate advocates for the organisation, spreading the word to a broader audience. This thesis addresses the second attempt to improve the user experience of mentorship platforms for mentors.

Re:Coded launched the Mentorship Platform – 'The Community Platform' – in response to changing demands. Initially powered by StellarUp, the organisation eventually switched to Mentorlink as a strategic move. However, The Community platform powered by StellarUp will be the exclusive subject of this thesis, serving as a case study and example of a mentorship platform. Nevertheless, the principles presented in this report have the potential to be implemented generically across a variety of mentorship platforms.

## 1.1 Research Question:

*How can Re:Coded optimise the user experience for their volunteer mentors so they can recommend the experience to their friends and family? How can the organisation use platforms like the Community Platform to achieve an excellent experience at scale, as they have a small team?*

## 1.2 Hypotheses:

**Hypothesis A:** If Re:Coded enhances the usability of The Community Platform by integrating feedback mechanisms and recognising mentor contributions, it will increase mentor motivation, satisfaction, and loyalty.

**Hypothesis B:** By patching The Community Platform to the unique needs of mentors and including a user-friendly experience, Re:Coded will elevate mentor satisfaction and promote more assertive advocacy for the organisation within the broader tech community.

# Chapter 2: Literature Review

Mentorship is an age-old tradition that bridges the gap between inexperience and competence in personal and professional growth. Sharing expertise and assisting someone in their professional development (Kalbfleisch, 2002). Historically established in ancient times, the practice has evolved and modified to reflect cultural and technical advancements. This adaptability is even more evident in the modern era, as technological innovations redefine how mentors and mentees communicate and engage. This literature study goes into the history of mentorship, examines its transformation with the introduction of technology, and throws light on the current issues and concerns in digital mentoring platforms. Through this exploration, we aim to gain a holistic understanding of the evolving landscape of mentorship in the digital age.

## 2.1 A Historical Overview:

Mentorship has a comprehensive background that may be traced back to ancient times. It is a matter of an experienced individual assisting someone less experienced (Kalbfleisch, 2002). This concept derives from an old Greek story in which Odysseus picked Mentor to assist and instruct his son, Telemachus (Young & Wright, 2001; Clever, 2000).

Mentorship has been more popular in educational institutions and the business sector in the modern era. When considering areas of professional achievement, Provencher, in a study in 2009, discovered that mentorship is the key to success in a competitive corporate environment where it is hard to succeed by playing by the rules. The need for a guiding hand grew with the beginning of industrialisation and complicated work positions. This move meant that the new generation of employees would be well-prepared for the rigours of their jobs. Mentorship practises evolved to meet the demands of particular professions and specialities effectively as industries and sectors diversified, as found in a study by Weijden in 2014.

## 2.2 Digital Transformation in Mentorship:

In the past, Mentorship used to take place primarily in closed communities, schools, or workplaces where mentors and mentees interacted face to face (Fowler & O'Gorman, 2005). These in-person interactions were critical in developing a solid trust, respect, and learning atmosphere between the two (Pelletier et al., 2020).

However, time has changed. Technology began to play an important role, especially in the second half of the 20th century. This change resulted in a shift in mentorship practices. Thanks to computers and the internet, people began mentoring across digital platforms, ranging from basic messaging applications to robust AI-driven systems (Muller, 2009).

As the internet became more widely available, online mentoring grew significantly. This shift meant that someone in Europe, for example, might obtain counsel from an expert in America without having to fly. Anyone, wherever, can study and get instruction by employing a computer. This approach provided opportunities for many worldwide who wanted to learn and progress.

However, this new online mentoring technique had its challenges. Indeed, it was handy, but there were concerns: Can an online mentor truly comprehend and assist as well as a face-to-face mentor? Today, research or review focuses on online mentoring, including its advantages, how it has altered things, and the challenges it may bring in today's tech-driven society (Heinonen et al., 2019).

## 2.3 Challenges in Digital Mentorship:

Although digital platforms provide unique options for mentorship, they also provide their challenges. One essential worry raised by Briscoe in 2019 is that many of the access restrictions that limit the success of traditional, face-to-face mentorships are eliminated by the virtual method. This method implies that the link between mentor and mentee may sometimes be more surface-level, missing the closeness and depth typically anticipated from mentorship.

Further investigation is required to understand the complexities of Briscoe's argument. Because of their nature, digital platforms prioritise accessibility and ease. While this ease of connection is advantageous in many ways, it can also lead to more temporary relationships. Unlike conventional mentorships, where both sides generally commit time and effort to schedule face-to-face encounters, digital platforms may mistakenly encourage short check-ins without substantial talk. In different mentorship relationships, digital ones are believed to not replicate on-to-on conversations (Heinonen, 2019).

## 2.4 Mentorship in Tech Education:

Mentorship has long been considered essential to the tech education environment (Arnesson & Albinsson, 2017). Because technology constantly evolves, integrating professional coaching with technology is a possible framework for modifying and sharing evidence-based treatments, ensuring students are aligned with industry expectations (Taylor et al., 2008). In this setting, mentoring links academic knowledge and real-world application, frequently resulting in enhanced student outcomes. For example, organisations such as Re:Coded have emphasised this significance by attempting to improve the digital skills of aspiring tech leaders through personalised mentorship.

While this thesis focuses on Re:Coded's mentorship platform, it is essential to note that the IT sector has experienced increased mentorship activities. Many technology educational institutions have made mentorship a fundamental component, recognising its tremendous influence on student preparation and adaptation in a changing technological world. The usefulness of such programmes has been debated within the business, emphasising the importance of constantly refining and adapting mentorship practises to suit changing educational and industrial demands, as Earnshaw mentioned in a study in 1995.

## 2.5 User Experience and Customisation:

The rise of digital mentoring platforms has brought out the importance of user experience. In digital media, UX comprises more than just the look and feel; it also includes the user's whole experience while engaging with a site (Wiryawan, 2011). This aim has everything from the first login and platform navigation to the way the material is displayed and how easily mentors and mentees may interact. While digital mentorship platforms are often discussed in the academic community, improving the UX of these platforms still needs to be explored, which is the primary focus of this study.

# Chapter 3: Methodology

This research used a User-Centred Design (UCD) approach to improve the user experience on the mentorship platforms. Adopting Stanford University's design thinking process. the research journey was meticulously divided into five pivotal stages, beginning with an in-depth understanding of user behaviours, needs, and challenges. The study progressed through a sequence of data analysis to find and define the main problems, design ideations for each case study, prototype each one of the case studies based on the findings from different methods, and finally, test the prototype with an expert.

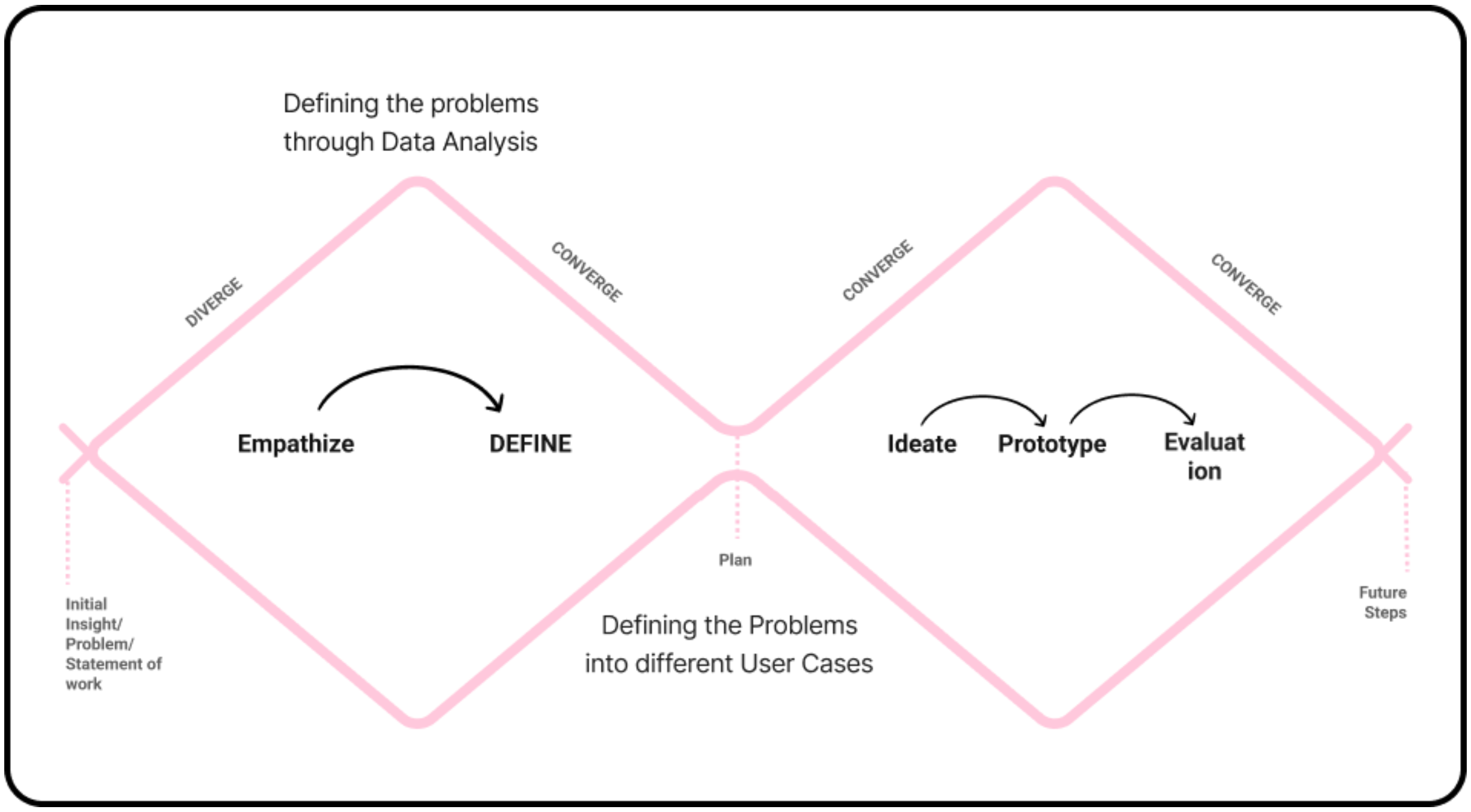

## 3.1 Research Philosophy

Understanding how people think and behave is critical in user experience design. As a result, our research employs a constructivist approach (Magoon, 1977). Proper understanding comes from seeing how people interact and acknowledging that everyone's experiences are shaped by their unique perspective.

Because there are so many aspects to consider in UX, we must appreciate and understand the various user experiences. Some strategies, such as quantitative methods, may need to provide a comprehensive picture of how users engage with digital technologies. This is why we picked qualitative research methods that were complementary to the strategy of our study. We divided the chosen methods into two groups based on our research questions and hypothesis. The first group mainly focused on understanding the user needs as mentors, while the other group concentrated on evaluating the Community Platform to understand the pain points of the current users.

## 3.2 Participants:

For this project, we focus exclusively on the mentor's aspect of the mentorship platforms. Consequently, the participants included in this study were mentors, educators, and senior professionals from the tech field. They held various positions ranging from UX designers and software developers to CEOs. The diversity in roles and locations was intentional, as we aimed to emulate the genuine mentor profile of Re:Coded, who came with a different range of positions from different locations around the globe and various levels of the mentoring experience.

## 3.3 Research Methods Employed to Identify Pain Points:

### 3.3.1 Heuristics Evaluation:

Heuristics evaluation is essential to user research methodology (Bertini et al., 2009). This evaluation approach allows specialists to analyse a user interface independently against a set of determined heuristics or best practice criteria. The key benefit of heuristic assessment is its capacity to quickly and cost-effectively identify possible usability flaws, which frequently translate into user pain points (Zhang et al., 2003).

In our research, the Heuristic Evaluation served as an essential initial tool. It facilitated our understanding of the platform's functioning and allowed us to identify systemic issues. This approach further guided the refinement of our research questions. Consequently, the insights from the Heuristic Evaluation informed our selection of subsequent research methods.

**Data Analysis:**

For data analysis, we kept to the heuristic evaluation template the Nelson and Norman Group provided in 2023, categorising the identified problems into ten groups, each corresponding to a specific rule.

**3.3.2 Usability Testing:**

Another primary approach in our study is usability testing, which involves real users interacting with the interface to measure its efficacy, efficiency, and overall satisfaction. This hands-on method provides direct insights into genuine user experiences, allowing researchers to find pain points and effective, intuitive design (Fairbanks & Caplan, 2004).
We conducted usability testing in the lab; a moderator joined each experiment to answer the testers' concerns and observe their interactions. During the usability testing phase, each of the eight participants - who were mentors or potential mentors in the tech field - was asked to complete a range of activities (see Appendix A). This approach allowed us to analyse their interaction patterns with the community platform and gather crucial work performance and engagement data. In our multidimensional approach to usability testing, we carefully evaluated the time required to complete each activity and diligently tracked users' success or failure in completing the given tasks.

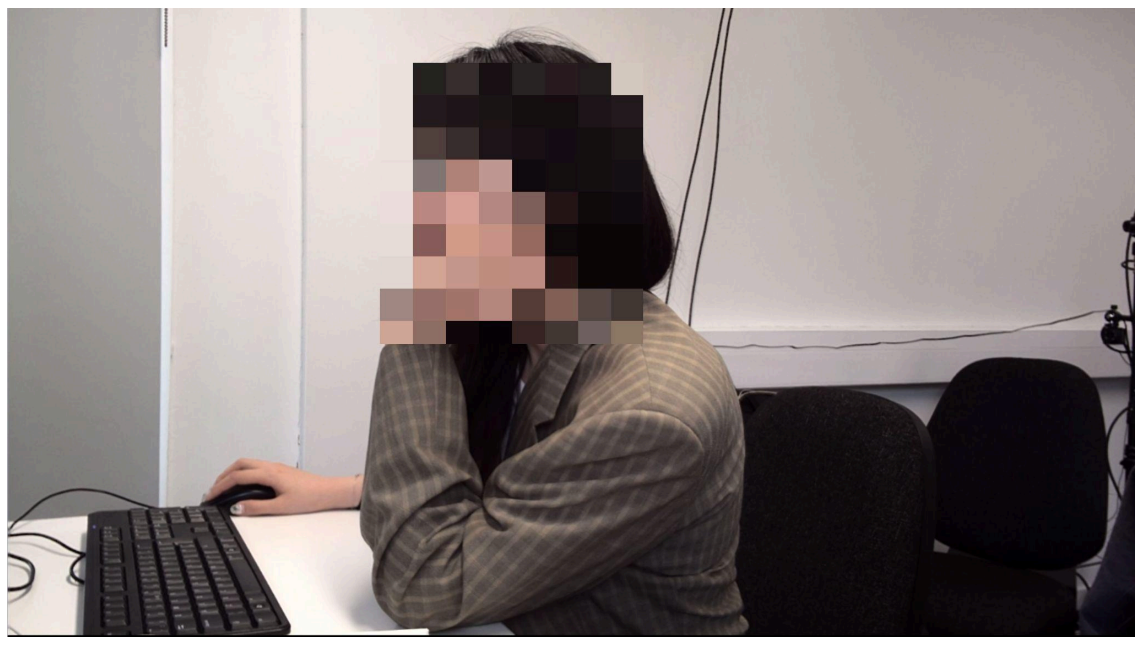
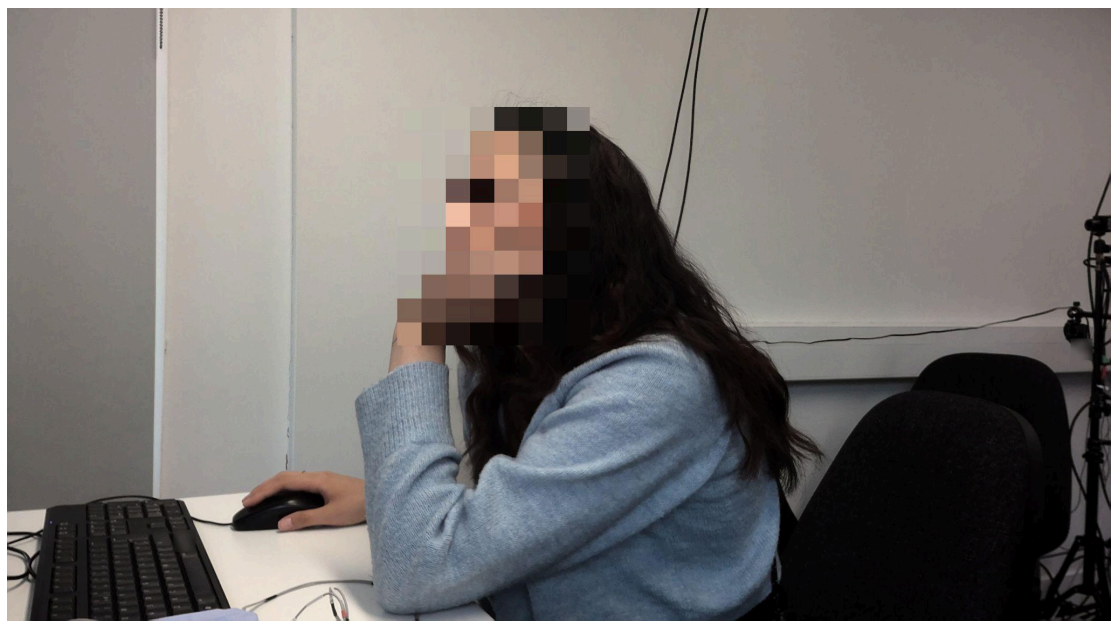

Image 1: Snapshots of the atmosphere during the usability testing.

**Data Analysis:**

In the data analysis section, we created a template for each tester that included the status of each task, comments from the users, and the amount of time required for each task. The main aim is to find any patterns in each task.

### 3.3.3 Eye Gaze Tracking:

As part of our usability testing, we used the Tobii Eye-tracking system to better understand the user behaviours of our eight testers while using the community platform from Re:Coded. The following stage of our research required extensive visual data processing, which included cleansing the data and removing any anomalous or incongruent readings.

**Data Analysis:**

The resulting cleaned data was then used to construct a heatmap, which was accomplished by superimposing gaze points onto a graphical representation of the stimulation to which the individuals were exposed. A gradient of warmer colours was used to emphasise regions within these incentives that grabbed more significant concentrations of the individuals' gaze, indicating increased times of visual engagement (Hillaire et al., 2012). We used our software's inherent capabilities to generate these heatmaps, using the built-in functionality developed for this specific purpose.

## 3.4 Research Methods Employed to Identify User Needs:

### 3.4.1 Interviews with Mentors Beyond Re:Coded:

In our study, we used user interviews to get in-depth insights into the participants' experiences, user needs, and points of view. These semi-structured interviews provided flexibility, allowing participants to openly express their opinions while ensuring the study objectives were met (Horton, 2012). All of our interviews were conducted online using both Zoom and Google Meets. The questions we asked were designed to generate rich, qualitative material, which was then transcribed and thematically analysed - See Appendix B for the Interview Questions.

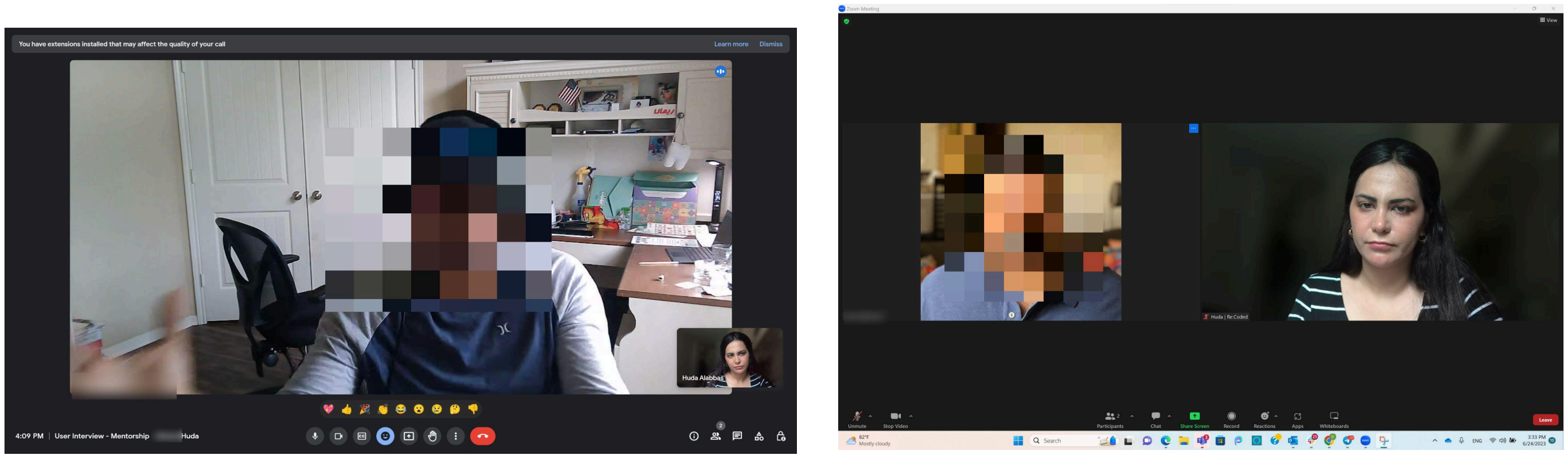

Image 2: Snapshots from the User interview through Zoom and Google Meets

The following step included a detailed data processing workflow after the precise execution of eight detailed interviews with notable mentors worldwide- reflecting the diverse community of mentors Re:Coded has. We began by converting the audio recordings into written transcripts, a procedure critical for preserving the reliability and richness of the material gathered during the talks.

**Data Analysis:**

Following the preparation of the transcripts, it was critical to edit the raw data to remove any potential discrepancies or redundancies. This cleansing procedure was critical in ensuring that only the most trustworthy and relevant information was kept for further examination.

Following the data-improving process, a compilation of significant insights was methodically extracted from each interview. This stage entailed a thorough review of the transcriptions and the extraction of significant notes, an undertaking that helped emphasise each mentor's unique insights and experience, and an affinity diagram was conducted to make a thematic group of data.

### 3.4.2 Focus Groups with Re:Coded Mentors:

We intended to employ study groups with current Re:Coded mentors as an additional data source. A series of interview questions was developed, focusing on their interactions with the platforms (see Appendix B). We used a variety of outreach strategies to recruit these mentors for the study groups, including sharing a survey inside the Community Platform and asking the mentors if they are interested in joining a focus group study. Unfortunately, no one responded to our invites, forcing us to drop out of this method.

### 3.5 Expert Review for Evaluation:

An expert review was conducted online with a UX designer and Front-end developer with over seven years of experience building and analysing digital platforms to assess the proposed solutions objectively. This UX professional now serves as a mentor in official and informal roles and is one of the interviewees during our project's research phase, providing them with relevant insights into the user viewpoint. A summary of the primary challenges found surrounding navigation, onboarding, and digital tool integration was provided to the expert. They were then shown the recommended solutions and high-fidelity prototypes for each area for feedback.

# Chapter 4: Findings

## 4.1 Data Analysis and Results:

### 4.1.1 Heuristics Evaluation:

Our heuristic evaluation showed various usability issues corresponding to Nielsen's interaction design principles. According to the idea of system visibility, the platform notably lacked suitable system notifications, particularly when users were paired with trainees or completed assignments. Furthermore, the distinction wall in the Cohort section needed to be clarified. The confusing usage of grey text and the unclear location of the 'Back' button, which pertains to consistency and standards, confused consumers about active vs. inactive alternatives. This uncertainty also hinted at some problems in error prevention.

Other problems occurred in user control and freedom since users could not remove submitted files and suffered navigation difficulties with the 'Back' button. The platform is confusing 'Mark as Done' and 'Pulse' capabilities, and the platform's uncertain difference between the Community Wall and Forum all came under the need for more consistency and standards. The default profile image's appearance, which looked to users as non-intuitive, generated concerns about Recognition rather than recall, as well as Aesthetic and simple design. While our review mainly focused on these aspects, it is important to note that heuristic evaluations can be subjective, with the opportunity for interpretation depending on more profound insights into the platform and user interactions. We have selected the most feasible and common problems that could be applied on all the mentorship platforms to be tested during the usability and Eye Gaze Tracking testing.

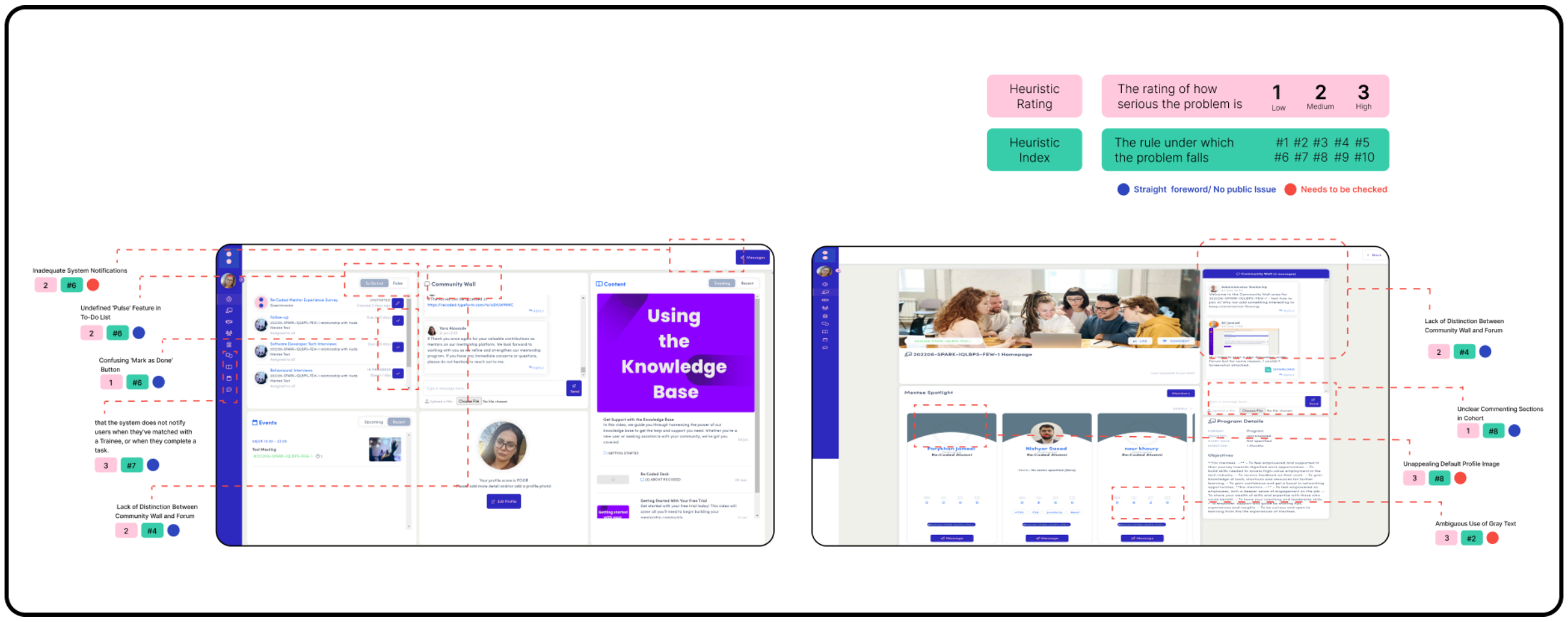

Image 3: Heuristic Evaluations on the Community Platform

### 4.1.2 Usability Testing

During our user testing, we discovered that four of the eight participants had difficulty navigating "the mentee's feature", which is essential to the functioning of our platform. This finding may highlight underlying problems with our user interface design's intuitive navigability. Furthermore, some participants needed to understand the platform's standards for creating a mentor-mentee connection. This uncertainty suggests a need for more precise instructions or orientation materials. Notably, this tendency coincides with another finding: the participants had yet to gain prior experience with mentorship systems, which will be discussed later in this chapter.

This emphasises the need for our application to adopt a more user-centric design and provide detailed advice, especially given the possibility that a significant portion of our user demography may be new to mentoring platforms. - See Appendix C for the rest of the tables -

| Tasks | Tester 1 | Tester 2 | Tester 3 | Tester 4 | Tester 5 | Tester 6 | Tester 7 | Tester 8 |
|---|---|---|---|---|---|---|---|---|
| Opining the application and send a message to a specific account | ~2.30m | 1.25m | 0.35m | ~1m | 0.45m | 0.30m | ~2.42m | ~2.30m |
| As a mentor, find the mentee you are matched with | 1.3m | 3m | 1.3m | 0.15m | 0.46m | X | 0.19m | 0.35m |
| Find "Manage your relationship" page and hand book | ~1m | 0.50m | X | 2.25m | X | X | X | 0.50m |
| Book a meeting with your mentee. | 0.50m | 3.30m | 1.40m | 1.17m | 1.22m | X | 1.17m | ~2m |
| Add a new skill to the list of your skills | 0.40m | 0.37m | 0.33m | 0.55m | ~1m | 0.16m | 0.40m | 0.26m |
| Find "Who is Re:Coded" article. | ~1m | 0.20m | 0.42m | 0.20m | X | 0.20m | 0.33m | 0.40m |
| oin the meeting you have created earlier. | ~1m | ~1m | ~1m | ~1.50m | 2.20m | X | 1.16m | 0.30m |

Image 4: Highlighted insights from the usability testing

### 4.1.3 Interviews:

From the insights we gained from the affinity diagram, we can analyse the takeaways from the points:

- **Formal vs. Informal Mentorship Approaches:** One of the most notable findings from the interviews was the participants' lack of use of mentorship platforms in general and organised mentorship approaches, which are usually adopted by these platforms. Instead of formalised processes, seven of the respondents preferred a more casual and informal approach to mentoring. This finding was demonstrated by their penchant for frequent conversations and providing advice sessions as needed. Such insights highlight a possible market emptiness in which formalised mentorship procedures find relevance and usefulness.

- **User-Centricity in Mentorship Platforms:** Another frequent topic in interviewee replies concerned the mentorship platform's target audience. A majority of six respondents said that these sites must primarily serve mentees. This insight emphasises the need to design these platforms to highlight the mentees' activities, milestones, and progress. Considering the motivating factors for mentors, there is a possibility of incorporating a reward mechanism. Seven respondents noted that their desire to mentor was inextricably linked to their willingness to give back to the community, consistent with Re:Coded's highest objective. This factor should be critical in determining platform design and functionality.

- **Digital Tools in Mentoring:** Understanding mentors' tools to optimise their duties might help shape future mentoring platforms. According to the interviews, five respondents used various digital technologies to monitor and improve their mentorship commitments. Google Calendar, Trello, ClickUp, Figma, and Mural were frequently mentioned. Their principal role is task management, which allows efficient interactions between mentors and mentees. This finding might indicate a more targeted and personalised feature set in future mentorship systems, with integrated task management modules taking precedence.

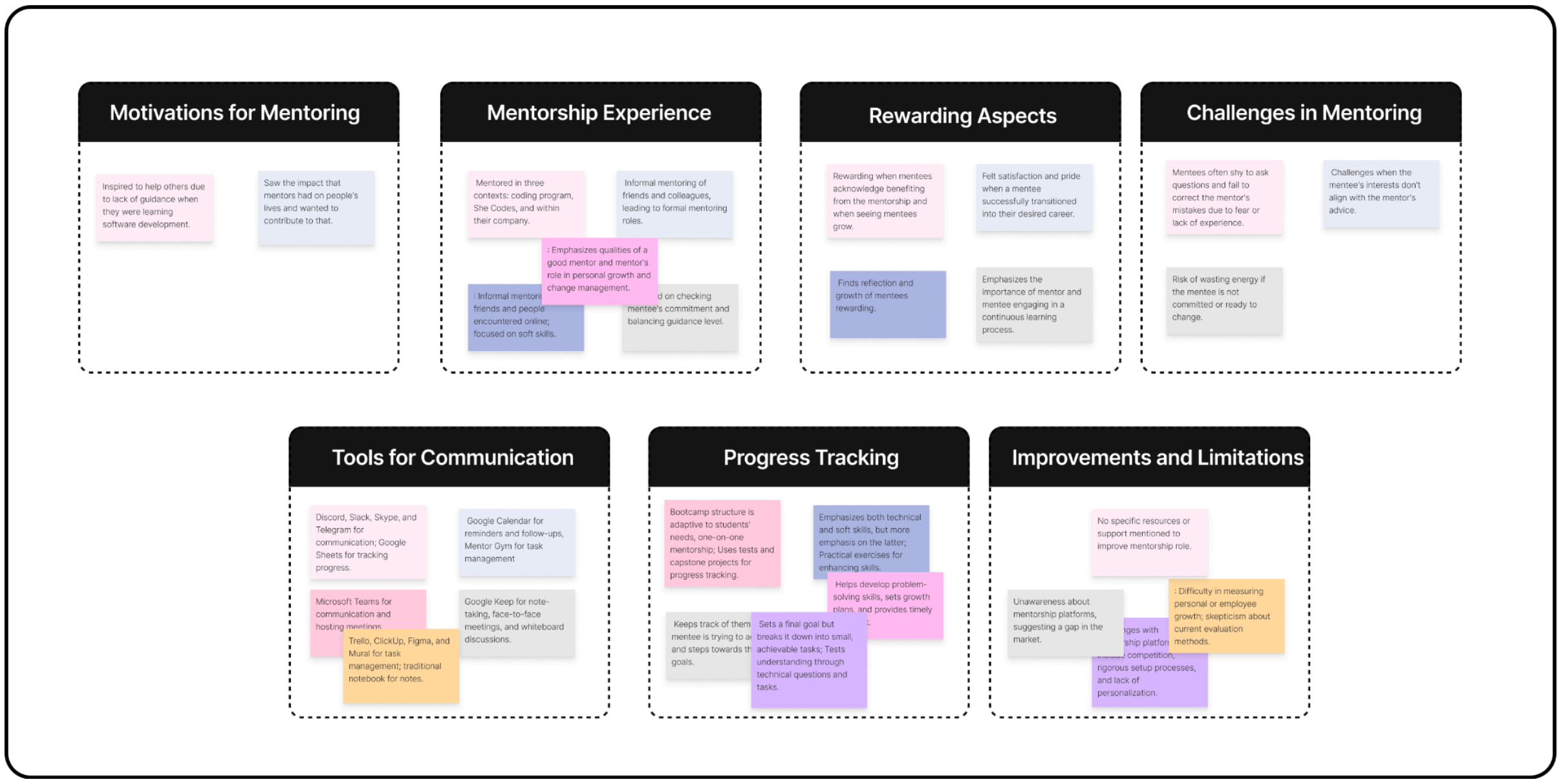

Image 5: Affinity Diagram from the User Interviews

### 4.1.4 Eye Tracking and Heatmaps:

During the analysis of the heatmap, we got a lot of exciting data; here is a snap of the most significant findings:

In image 6, the heatmap visually represents the specific areas on the Content page where visitors focus their attention. The majority of the engagement is concentrated on the content's title. In contrast, the accompanying file garners significantly less interest. This design decision aligns with insights from our usability testing, where users often needed help to locate shareable content and resources.

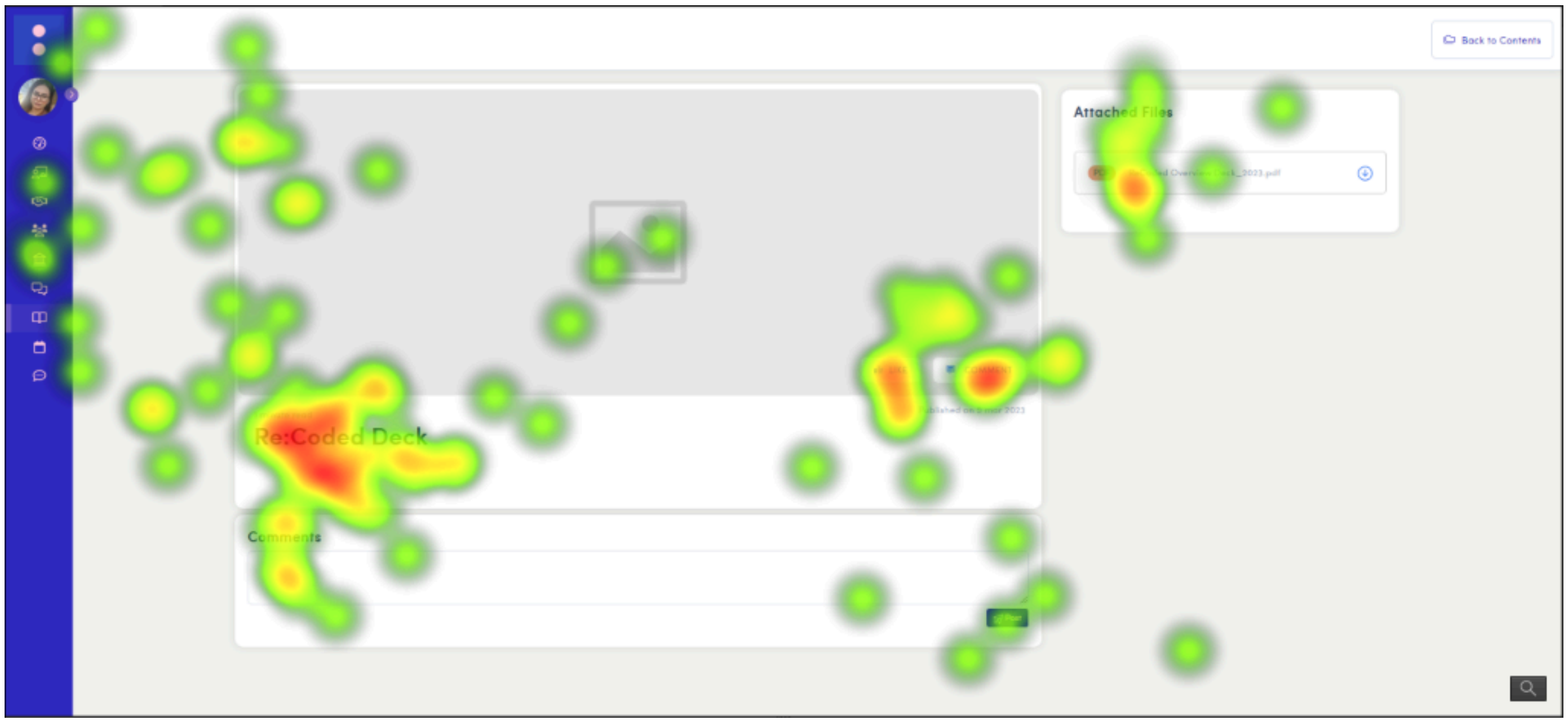

Image 6: The Heatmap of the Content Page

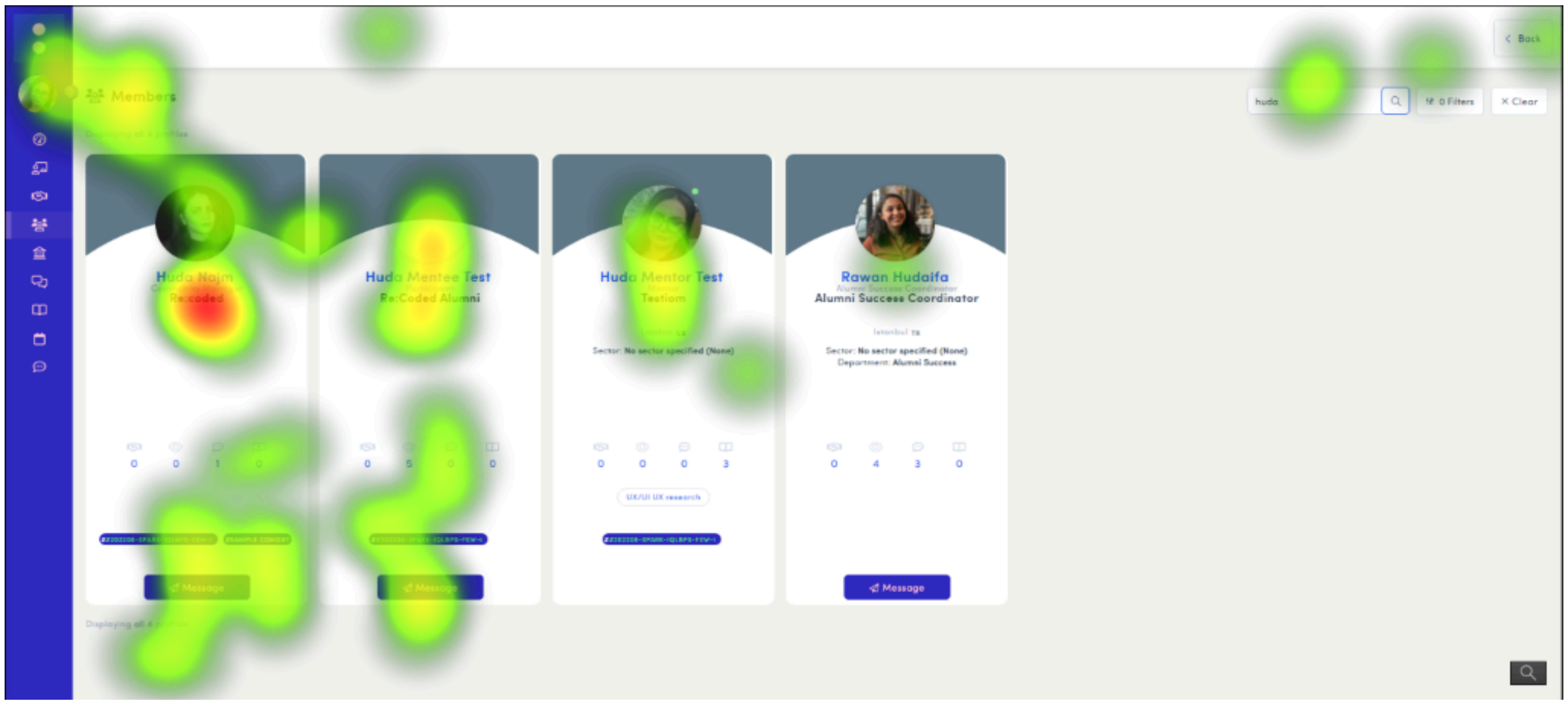

Image 7: The Heatmap of the platform users

in Image 7 focuses on usernames and their associated remarks within the user cards. Conversely, users need to pay more attention to other status indicators within the cards. This observation aligns seamlessly with our heuristics evaluation results, suggesting that specific symbols are not prominently displayed and lack sufficient importance.

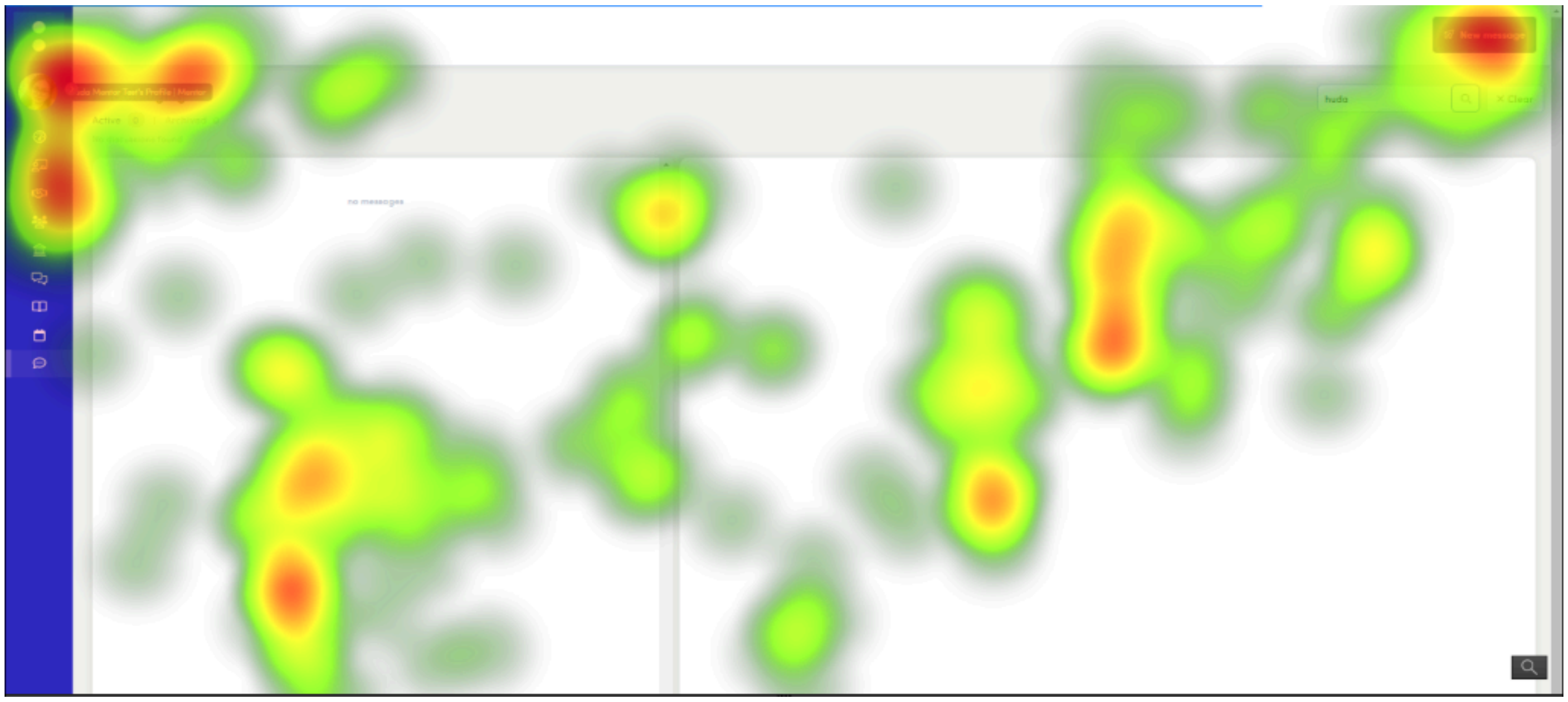

Image 8: The Heatmap of Conversations Page

The visualisation in image 8 underlines a fundamental ambiguity within the sending message feature. When prompted to send a message, users encounter an unintuitive UX, causing uncertainty about the messaging process. Consequently, they second-guess their navigational choices, oscillating between various pages. Their frequent peeks at the navigation bar further accentuate this sense of disorientation.

## 4.2 Synthesis of Findings:

Our comprehensive analysis found numerous critical aspects of the mentorship platforms' design and performance. Here is a summarised overview:

- **Navigation:** Confusion is a frequent topic in navigation. There is a strong desire to simplify and explain the platform's navigation paradigm, from issues distinguishing between primary platform functions to difficulties executing fundamental actions like finding the mentees. This improves user autonomy and will likely minimise recurring page switching, as seen by users' frequent usage of the navigation bar.

- **Onboarding:** Since many users are new to mentorship platforms, supporting a user-centric design strategy is critical. This would ideally include an intuitive onboarding procedure combined with straightforward instruction, ensuring users, particularly newbies, easily find their footing.

  According to our findings from the user interviews and user testing, many mentors who are significant actors in these platforms must be made aware of the subtleties of mentorship platforms.

  Onboarding is more than just showing a new user around a platform; it is about making them feel at ease. How the onboarding process is designed for users unfamiliar with specific software can determine whether they remain or depart fast. Onboarding serves numerous purposes, especially for mentorship platforms that need to be more familiar with formal mentoring. It instructs users on the platform's design, features, and operation. Following that, users will be encouraged to delve deeper and make more use of the site (Baker, 2019).

- **Digital Tools Integration:** Using the power of popular digital tools such as Google Calendar, Asana, and Calendly may improve the platform's usability and create a more seamless user experience by expediting tasks and engagements.

- **Interface and Recognition:** The requirement for a more intuitive interface is a recurring topic in the studies. The design of the default profile photograph needed to be clarified. The grey lettering obscured the contrast between active and inactive choices, worsened by the ambiguous location of the 'Back' button. Similarly, some confusion was observed in distinguishing between the Community Wall and the Forum and where to find the "mentee's feature." Ambiguities surrounding capabilities such as 'Mark as Done,' 'Pulse,' and the dividing wall in the Cohort section highlight the need for a more defined design approach. The heatmap data revealed that certain symbols were invisible, emphasising the need for visual simplicity and a design prioritising recognition over recall.

By addressing these synthesised findings, the mentorship platforms, generally and the community platform specifically, can become more user-friendly and robust.

As the first three findings are more general and could be applied to a wide range of mentorship platforms, This thesis will address only them as cases to find solutions for each. We have selected these 3 problems based on their generality rather than related to the Community Platform and the feasibility of applying these features. We have used a prioritisation matrix for this purpose.

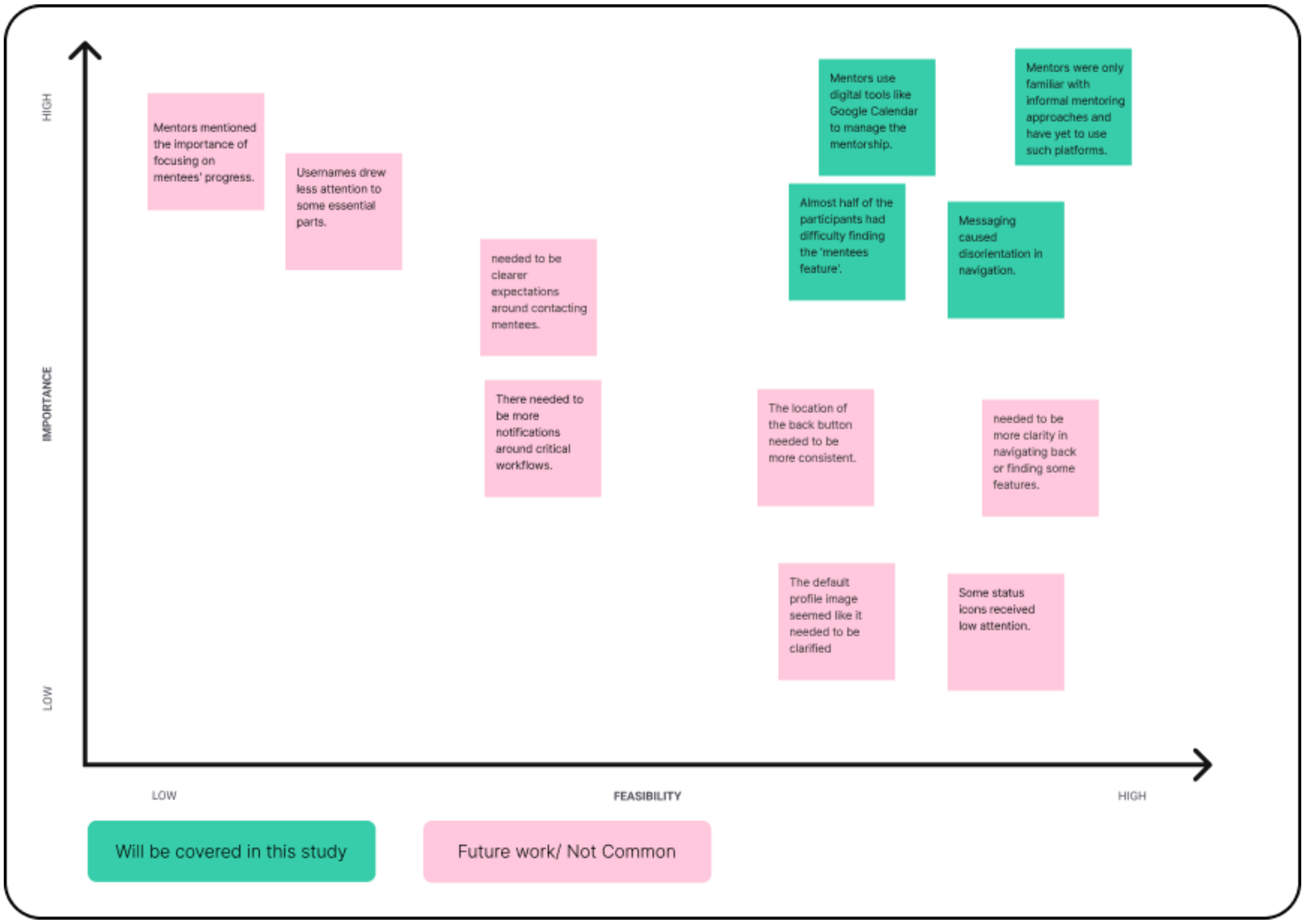

Image 9: prioritisation matrix of the findings

# Chapter 5: Suggestions and Solutions

## 5.1 Case One: Navigation

### 5.1.1 Problem Statement:

New community system users encounter challenges accessing specific features and navigating the platform efficiently. A more intuitive and well-structured navigation system is essential to enhance their user experience.

### 5.1.2 Used Method:

We selected an open-ended Card Sorting approach to solve this issue. We aimed to learn about participants' preferences for the platform's structure and their recommendations for naming the group titles for better hierarchy. We recruited the help of eight mentors for this activity. The study was held online using the 'UXweak' platform.

### 5.1.3 Findings and Proposed Solution:

In our analysis of the data received, and after considering the diverse features present in analogous platforms, we have strategised a division of the principal features within our application into four distinct groups.

The first group comprises "Messages," "My Calendar," "Events," and "To-Do List." Although we received several naming suggestions, such as "My Space" and "My Account," we have chosen the label "My Space" for its clarity and widespread familiarity.

The subsequent group encapsulates "My Mentee's Progress Bar" and "My Mentee To-Do List." We deliberated between the names "My Mentees" and "Mentees." Given the platform's design, where different mentors might supervise mentees, we adopted "My Mentees" to emphasise personal supervision.

Our third cluster integrates the list of mentees and mentors active on the platform. For this group, we have selected the title "Community" to reflect the interconnectedness of all members.
The fourth group amalgamates the "Community Wall" and "Learning Content." We have yet to receive specific naming suggestions for this group and will finalise its designation after further deliberation. Below, you can see the Similarity Matrix - which we depended on to analyse the data from the card sorting.

| | | | | | | | | | |
|---|---|---|---|---|---|---|---|---|---|
| My Mentee progre... | | | | | | | | | |
| 71 | My mentee to do list | | | | | | | | |
| 14 | 14 | The platform's me... | | | | | | | |
| 29 | 29 | 57 | The platform's me... | | | | | | |
| 14 | 14 | 0 | 0 | My calendrer | | | | | |
| 14 | 14 | 0 | 0 | 71 | To do list | | | | |
| 14 | 14 | 0 | 0 | 57 | 57 | Messages | | | |
| 0 | 0 | 14 | 14 | 43 | 29 | 14 | Events | | |
| 0 | 0 | 14 | 14 | 0 | 0 | 29 | 29 | Community wall ( ... | |
| 14 | 14 | 14 | 14 | 0 | 14 | 0 | 14 | 29 | Learning Content |

Image10: Similarity Matrix of the Card Sorting Activity

After analysing the data, we created an application map to outline the structure and hierarchy of the application.

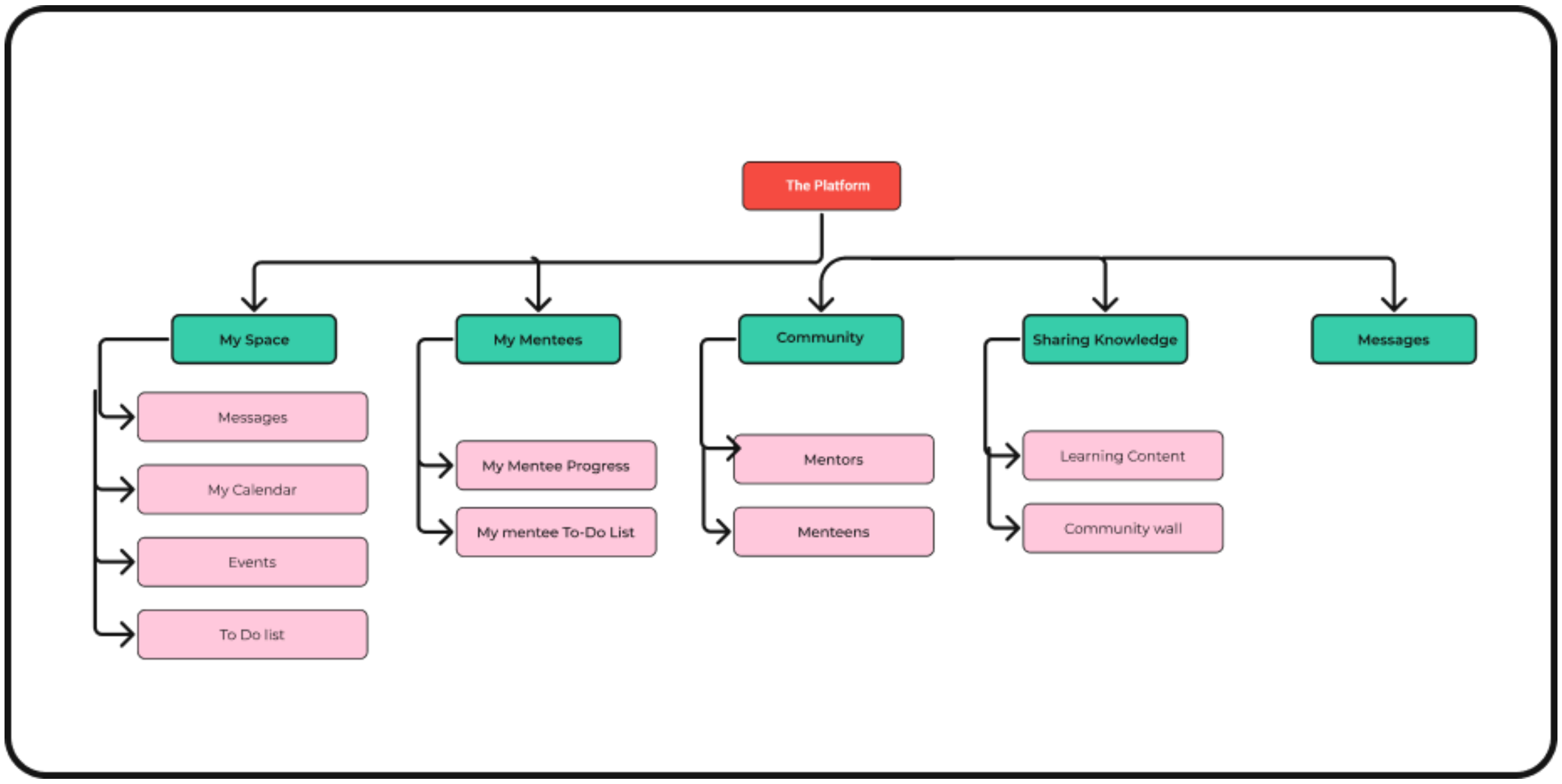

Image 11: The Suggested Ideal Application Map for Mentorship Platforms

**5.1.4 Proposed Prototype:**

The images below highlight each section's navigation bar. It mainly shows the hierarchy of the Sharing Wall page, which includes both the "Community wall" and the Learning consent based on the suggestions from the Card Sorting.

The prototype shows the "On the Platform" page, which includes all of the users of the platforms, including other mentors and all the mentees.

In addition, the prototype reflects the content of both "My Space" page, which includes the calendar and To-do List, as well as "My Mentees" sub-page, showing what the individual page for each mentee should look like.

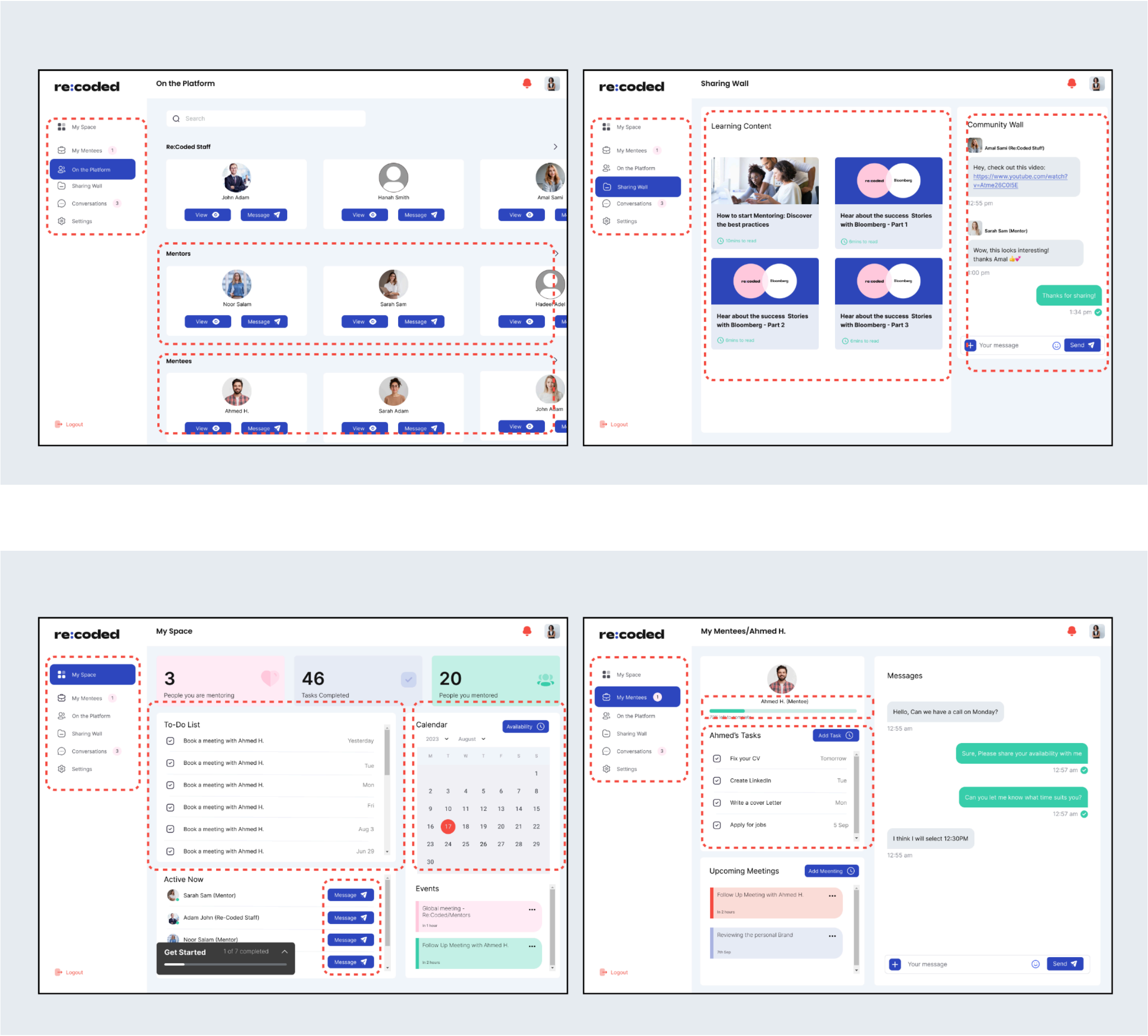

Image 12: High-fidelity prototype showing the navigation system

## 5.2 Case Two:  Onboarding and User Guidance:

### 5.2.1 Problem statement:

Mentors with no prior experience on mentorship platforms or in formal mentoring approaches require a comprehensive onboarding system and guidance to ensure seamless and efficient platform use.

### 5.2.2 Method Used:

We have relied on secondary research and the mentors' suggestions for this issue. Additionally, we have considered the organisation's requirements to identify the best practices for constructing a formal onboarding process for mentors.

### 5.2.3 Proposed solution:

Below are the Suggested Solutions for Mentor Onboarding and Engagement:

- Understanding Mentorship: Incorporating Educational Videos: The way mentors perceive the benefits of the mentorship process is closely tied to the methods employed in matching them with mentees. Research by Finkelstein Poteet (2007) suggests that the intricacies and dynamics of mentoring programs differ based on the involvement and methods utilised in the pairing process.
- Integrating educational videos can provide clarity about the mentorship process. When mentors are empowered to have an active role in the selection process, they have greater control over the relationship. This proactive involvement can subsequently lead to more benefits being extracted from the mentorship program, as suggested by Parise. (2008).
- Matching the Peers: Implementing Best Practices: The essence of a successful mentorship lies in the symbiotic relationship between mentors and mentees. The compatibility of this pairing significantly affects the outcome of the mentorship. It might be advantageous to give mentors the autonomy to choose their mentees.
- Individuals naturally tend to gravitate towards those they perceive as similar to themselves, leading to potentially more harmonious relationships. as suggested by Parise. (2008).
- Pick Your Availability: Empowering mentors with a system that allows them to mark their availability can streamline the mentor-mentee interaction process. Setting clear timeframes respects the mentors' schedules while ensuring that mentees can find a suitable slot.
- Review the CV of the Mentee: Before committing to a mentorship, mentors should access the CV or profile of potential mentees. This insight allows mentors to gauge the potential fit, understand the mentee's background, and tailor their guidance accordingly.
- Post-Initial Mentorship Session – Assigning the First Task: Setting direction and expectations is crucial after the first mentoring session. Assigning the first task to the mentee can act as a stepping stone, establishing a clear path for growth and learning in subsequent sessions.

Here is a flow chart showing the onboarding flow.

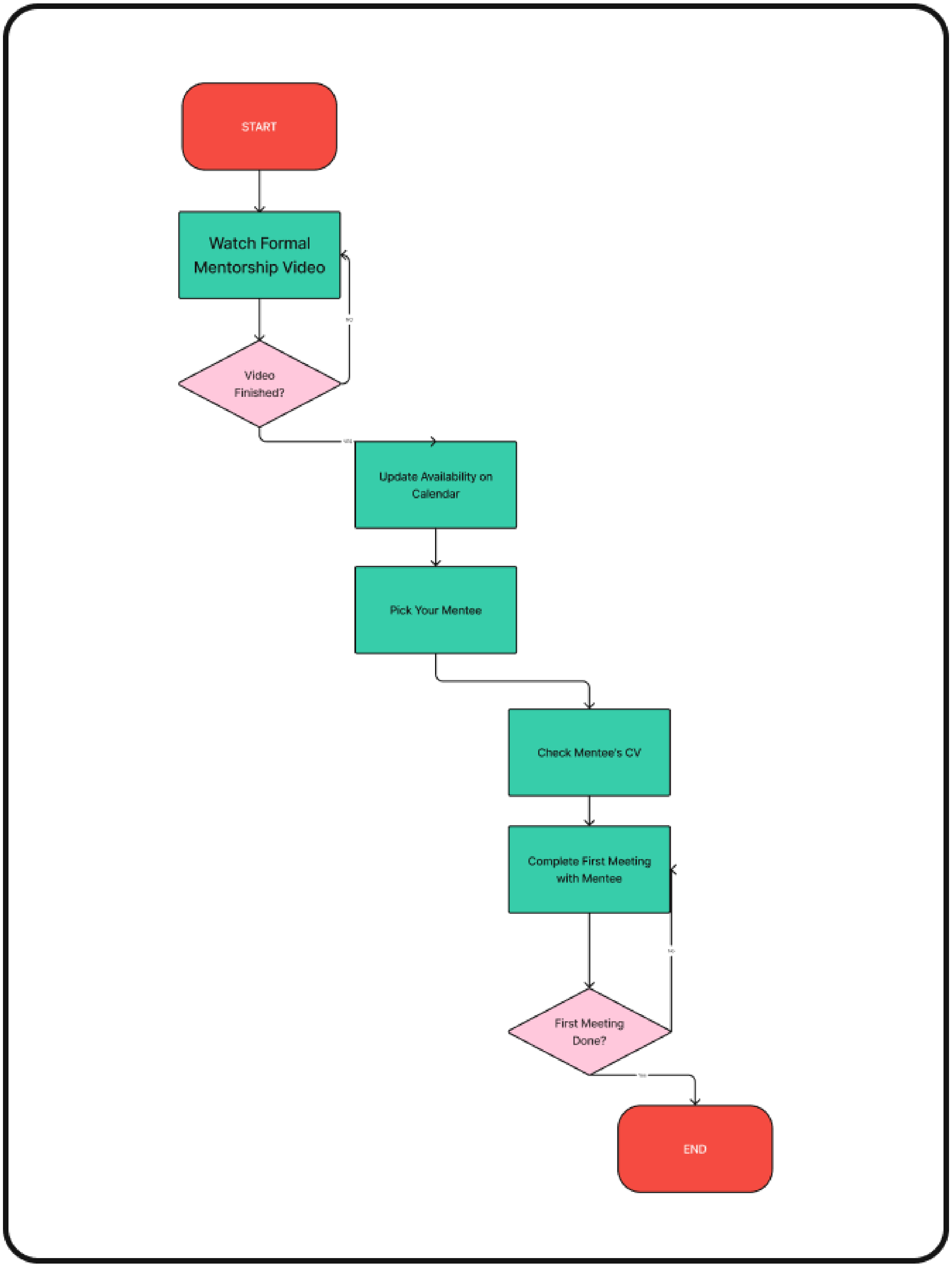

Image 13: High-fidelity prototype showing the onboarding system

### 5.2.4 Prototyping

This prototype features a card on the side outlining the essential steps mentors must undertake upon onboarding onto the system. This guidance is crucial for those new to the platform and for mentors who still need to acquire experience in formal mentoring. Additional instructions for completing each subsequent step are revealed once the mentor completes the preceding one to facilitate a smooth progression. This sequential revelation of instructions ensures a structured and guided approach for mentors, enhancing their onboarding experience and preparing them effectively for their mentoring role.

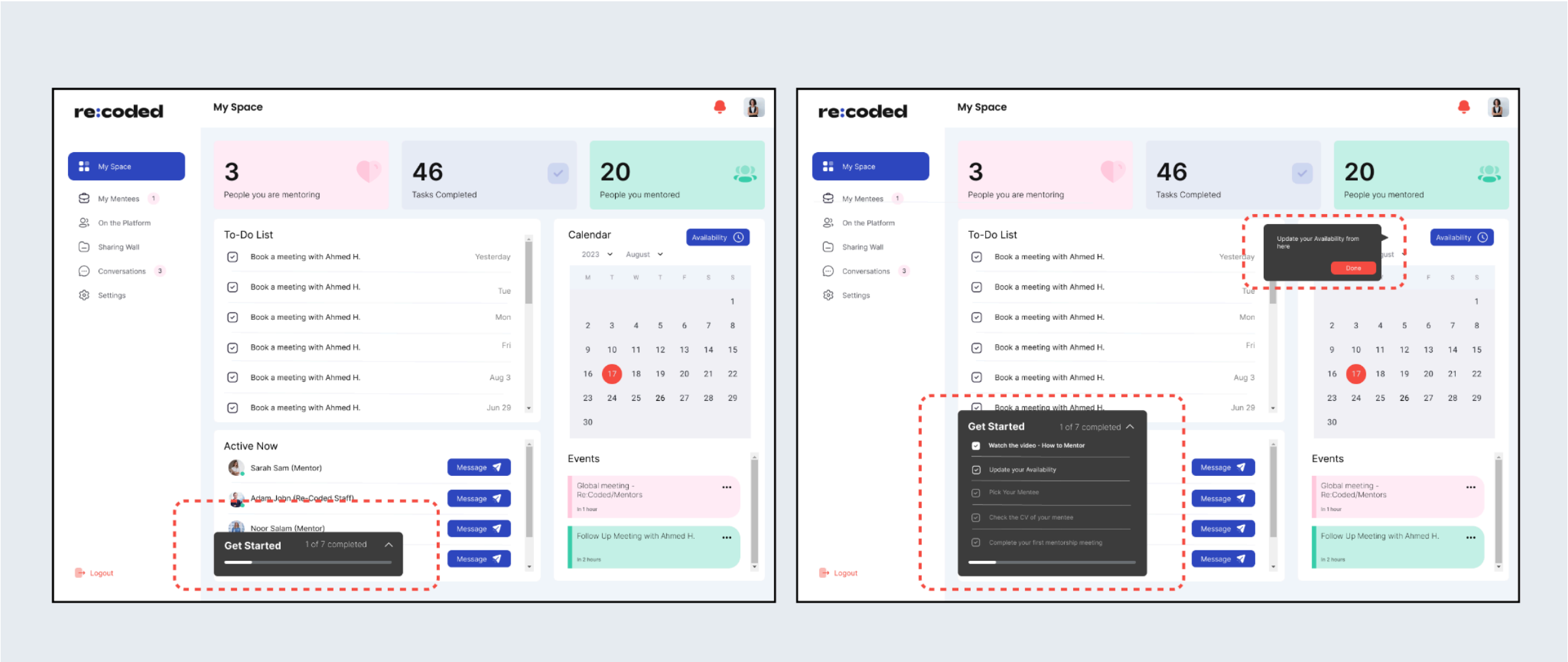

Image 14: High-fidelity prototype showing the onboarding system.

## 5.3 Case Three: Digital Tools Integration

### 5.3.1 Problem statement:

Our comprehensive research shows that mentors in informal settings consistently resort to alternative communication tools when interacting with their mentees. By integrating features and concepts from these preferred platforms, we have the potential to enhance the usability and adoption rates of our mentorship platforms.

### 5.3.2 Used Method and Proposed Solution:

We opted to evaluate our indirect competitors to address this specific challenge meticulously. We dedicated focused attention to each platform, discerning its most prominent features. The chosen features and suggestions for their integration into a mentorship platform are presented below.

**Asana**

| Feature | How it works | How to reflect on the mentorship platform |
|---|---|---|
| Task Management | Asana is renowned for its project and task management features. | In a mentorship context, mentors can set milestones or tasks for their mentees. This ensures that the mentee has clear goals to achieve before the next session. |
| Progress Tracking | On Asana, there is a feature of showing the progress of each feature. | Offer visual representations (like progress bars or charts) that show how much of the mentorship curriculum or milestones have been achieved. |

**Calendly**

| Feature | How it works | How to reflect on the mentorship platform |
|---|---|---|
| Availability | Calendly's strength is in showing availability based on one's calendar. | Implement a feature where mentors can sync their calendars to show their free slots, making it easier for mentees to book sessions without the back-and-forth communication. |
| Time Zone Adjustments | On Calendly, you can adjust your time zone. | If your mentorship platform caters to a global audience, ensure that the scheduling tool automatically adjusts for time zones. |

**Google Calendar**

| Feature | How it works | How to reflect on the mentorship platform |
|---|---|---|
| Scheduling | The Users of Google calendar have the ability to see the events of others in connection with, schedule, reschedule, or cancel sessions. | The mentors could have the ability to see the events their mentors has, schedule, reschedule, or cancel sessions with the mentee. |
| Reminders | Google Calendar provide a feature of reminding about the upcoming event. | Set up automated reminders for upcoming mentorship sessions by email. |

Image 15: Competitive analysis

### 5.3.3 Proposed prototype:

This prototype incorporates several critical features adopted from established platforms, enhancing its functionality and user-friendliness. An 'Availability Feature' has been integrated and modelled after the one found in Calendly, a widely used scheduling tool. This feature allows for efficient scheduling and management of availability. Additionally, a 'Tasks List' feature, inspired by the one in Asana, a popular task management tool, has been included to facilitate the organisation and tracking of tasks. Similarly, the prototype allows mentors to monitor their mentees' progress actively. This includes tracking their activities and managing meetings, with the functionality to add and edit meetings as necessary.

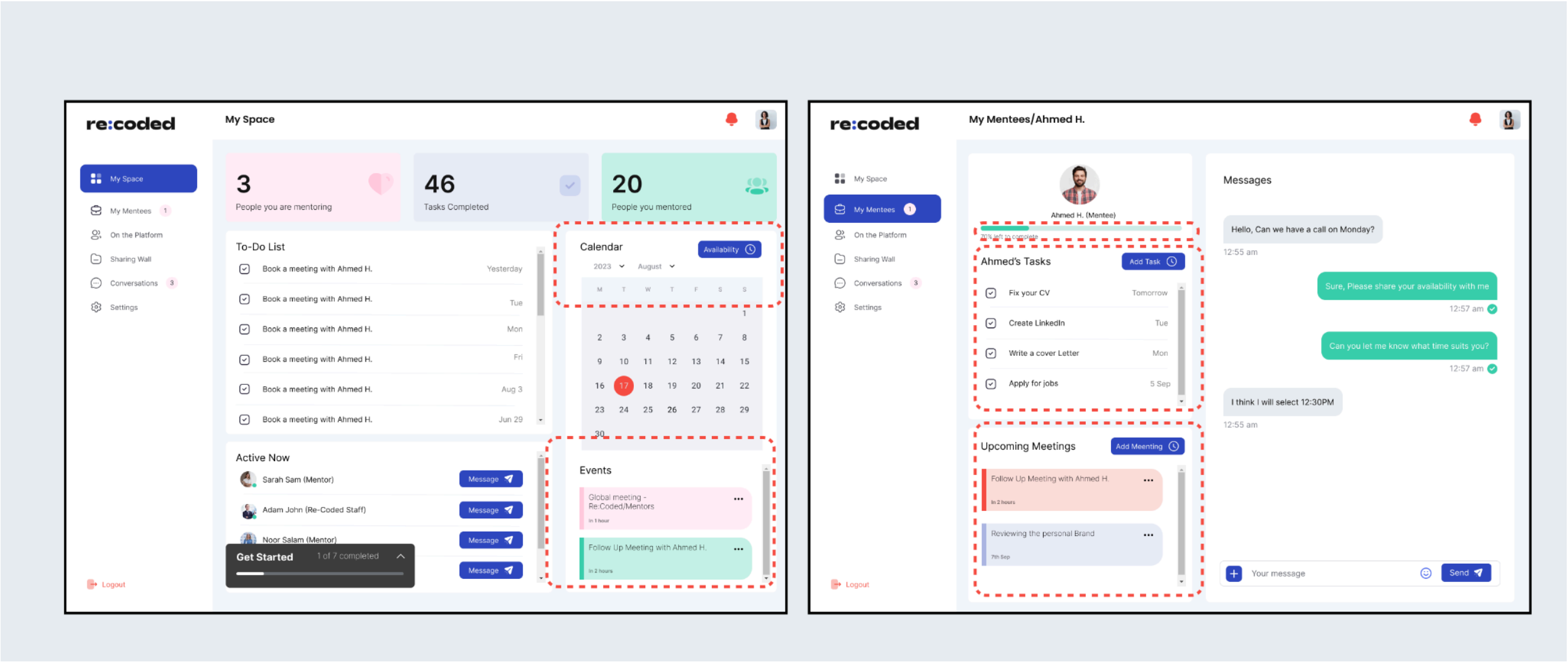

Image 16: High-fidelity prototype showing the integration of the tools.

# Chapter 6: Evaluation

During and after the session with the Expert, They provided feedback on each solution's feasibility and expected user impact. Regarding the navigation system, they felt the simplified sitemap and categories would make the platform structure more intuitive for users. However, they cautioned that consistent navigation patterns across all platform pages would be critical. On onboarding, they agreed that formalising the process should get new users up to speed more quickly. They suggested complementing the educational resources with in-app prompts and tooltips. They validated that linking mentors' existing tools would streamline workflows for digital tool integration but noted potential challenges in aligning different software systems. Overall, the expert deemed the solutions viable and user-focused while highlighting additional factors to consider during implementation, especially with the UI part. Their hands-on perspective as a designer and mentors lends credibility to the proposals, and all their feedback will be considered in the sound alteration.

# Chapter 7: Discussion

This research provides valuable insights into enhancing the user experience for mentors on digital mentorship platforms, though some limitations should be noted. The study was heavily focused on the Re:Coded Community Platform after their request to help them enhance their mentors' user experience. However, the findings are generalised to other platforms to fit any mentorship platform Re:Coded could consider in the future.

The sample of participants, while reasonably sized, came mainly from the tech sector. Including mentors from other fields could reveal more diverse perspectives and needs. Similarly, the majority needed more prior experience with formal mentorship platforms. Surveying more seasoned platform users may uncover more profound UX challenges.

Despite these constraints, the mixed-methods approach provided a well-rounded assessment of mentor needs and platform issues. The solutions presented aim to address the core problems identified directly. However, testing their implementation through prototyping and user feedback would benefit future work.

This thesis centred on the mentor experience, but enhancing usability for mentors and mentees is crucial. Comparative studies on the mentee's UX could uncover areas of balance in platform design. Beyond usability factors, examining how platform experiences shape mentor-mentee relationship development represents an impactful direction for further exploration.

However, this project had some limitations. While this research aimed to generalise findings across platforms, it focused heavily on a single organisation's platform. Analysing a more comprehensive range of platforms could surface a greater diversity of user needs. Additionally, the participant sample was drawn primarily from technology fields, including mentors from other industries, which may provide more varied perspectives. It was also recognised that many participants needed more prior formal mentorship platform experience. Consulting more seasoned platform users could highlight different UX challenges to address. This project centred on optimising the mentor experience, but assessing the mentee's UX in parallel is essential for balancing design tradeoffs.

This project makes substantial headway in highlighting and tackling key UX obstacles for mentors on digital platforms. However, additional research and testing will be instrumental in refining the design principles and recommendations that surfaced here. Overall, this thesis stresses the merits of participatory, user-focused design in creating platforms that successfully facilitate online mentoring.

Centering end-user perspectives promises more intuitive interfaces and connected mentoring communities.

## Chapter 8: Conclusion

This thesis highlights the value of human-centred design in creating digital platforms that foster positive user experiences and engagement, like digital mentoring. Through comprehensive qualitative research, core pain points were uncovered regarding the user experience of mentors on Re:Coded’s Community Platform. Issues identified around navigation, onboarding, and integrating external tools inspired clear opportunities for targeted UX improvements.

The solutions proposed, including an intuitive navigation structure, formalised onboarding, and linking existing productivity tools, aim to directly address mentor needs based on insights derived from real-world users. While focusing on one organisation’s platform, the research methodology and findings may inform design decisions for mentorship platforms across contexts.

By embracing participatory methods and placing user perspectives at the heart of the design process, this thesis models an approach to UX optimisation that could be applied broadly and specifically. Testing and iterating upon these solutions will be instrumental in the future in translating the concepts into measurable usability and user satisfaction improvements. However, the user-driven foundation established here highlights the human-centred thinking that promises to unlock the full potential of mentorship platforms.

In conclusion, The focus on elevating user experience ultimately serves Re:Coded’s greater mission of empowering new generations of tech professionals through mentorship. More intuitive design and frictionless interactions pave the way for more rewarding mentor-mentee relationships and meaningful impact at scale. In this sense, enhancing mentor UX has ripple effects that touch the entire community. This thesis provides philosophical and practical guidance for putting people first in pursuing that goal.

## Appendix A

**Usability Testing Questions:**

1- Open the application and message a specific account.

2- As a mentor, find the mentee you are matched with.

3- Find the “Manage your relationship” page and handbook.

4- Book a meeting with your mentee.

5- Add a new skill to the list of your skills.

6- Find the “Who is Re:Coded” article.

7- Join the meeting you have created earlier.

## Appendix B

**Interview Questions:**

1. Briefly describe your mentoring experience so far.
2. What prompted you to become a mentor?
3. Can you share a mentorship experience that you found particularly rewarding? What made it so?
4. Have you encountered any challenges in your role as a mentor? If so, could you describe them?
5. Have you used any specific mentorship platforms? If so, which ones, and what was your experience like?
6. What features or tools were most helpful on these platforms?
7. Were there any aspects of the platforms you used that you felt could be improved? If so, what are they?
8. How do you typically manage your mentorship activities without using a mentorship platform?
9. What would an ideal mentorship platform look like? What features would it have?
10. How do you communicate with your mentees? Is there a specific mode of communication you prefer?
11. How do you track the progress of your mentees? What tools or methods do you use?
12. What resources or support would make your role as a mentor easier or more effective?

**Focus Group Questions:**

1. Can you describe your overall experience with the StellarUp platform?
2. What features or tools do you use most frequently in the StellarUp platform?
3. How would you describe your experience with these frequently used features?
4. Can you share a positive experience you had while using the StellarUp platform?
5. Can you share a challenging or negative experience you had while using the StellarUp platform? How did it impact your mentoring process?
6. What are the most vital aspects of the StellarUp platform?
7. In contrast, what areas of the StellarUp platform need improvement?
8. Could you describe an instance where you faced difficulty in navigating or operating a feature on the StellarUp platform?
9. How do you feel about the communication tools within the StellarUp platform? Are there any changes you suggest?

10. Can you tell me when you needed assistance with the platform? How was the support experience?
11. What additional features or improvements could enhance your experience as a mentor on the StellarUp platform?
12. Are there any features that could be more appropriate or more appropriate for the user experience?
13. Can you discuss your experience with the platform's compatibility across different devices (e.g., mobile, tablet, desktop)?
14. Based on your experience, how could the StellarUp platform better support the interaction between mentors and mentees?
15. How do you think the StellarUp platform impacts the overall mentorship experience at Re:Coded?

# Appendix C

## Usability Testing Tables for the eight participants

| Tasks | Tester 1 | Tester 2 | Tester 3 | Tester 4 | Tester 5 | Tester 6 | Tester 7 | Tester 8 |
|---|---|---|---|---|---|---|---|---|
| Opining the application and send a message to a specific account | ~2.30m | 1.25m | 0.35m | ~1m | 0.45m | 0.30m | ~2.42m | ~2.30m |
| As a mentor, find the mentee you are matched with. | 1.3m | 3m | 1.3m | 0.15m | 0.46m | ✕ | 0.19m | 0.35m |
| Find "Manage your relationship" page and hand book. | ~1m | 0.50m | ✕ | 2.25m | ✕ | ✕ | ✕ | 0.50m |
| Book a meeting with your mentee. | 0.50m | 3.30m | 1.40m | 1.17m | 1.22m | ✕ | 1.17m | ~2m |
| Add a new skill to the list of your skills | 0.40m | 0.37m | 0.33m | 0.55m | ~1m | 0.16m | 0.40m | 0.26m |
| Find "Who is Re:Coded" article. | ~1m | 0.20m | 0.42m | 0.20m | ✕ | 0.20m | 0.33m | 0.40m |
| oin the meeting you have created earlier. | ~1m | ~1m | ~1m | ~1.50m | 2.20m | ✕ | 1.16m | 0.30m |

## User Testing

User ID: **User #7**
Test date: **27.06.2023**

| Activity | Time | User's thoughts | Success/Fail |
|---|---|---|---|
| 1- Opining the application and send a message to a specific account | ~2.42m | | ✅ |
| 2- As a mentor, find the mentee you are matched with. | 0.19m | | ✅ |
| 3- Find "Manage your relationship" page and hand book. | 1.20m | | ❌ |
| 4- Book a meeting with your mentee. | 1.17m | | ✅ |
| 5- Add a new skill to the list of your skills | 0.40m | | ✅ |
| 6- Find "Who is Re:Coded" artical. | 0.33m | | ✅ |
| 7- Join the meeting you have created earlier. | 1.16m | The community wall kept distracting me | ✅ |

## User Testing

User ID: **User #8**
Test date: **27.06.2023**

| Activity | Time | User's thoughts | Success/Fail |
|---|---|---|---|
| 1- Opining the application and send a message to a specific account | ~2.30m | | ✅ |
| 2- As a mentor, find the mentee you are matched with. | 0.35m | | ✅ |
| 3- Find "Manage your relationship" page and hand book. | 0.50m | | ✅ |
| 4- Book a meeting with your mentee. | ~2m | | ✅ |
| 5- Add a new skill to the list of your skills | 0.26m | | ✅ |
| 6- Find "Who is Re:Coded" artical. | 0.40m | Could not understand the point of this | ✅ |
| 7- Join the meeting you have created earlier. | 0.30m | Kept checking the home page thinking that it would be there | ✅ |

## User Testing

User ID: **User #5**
Test date: **27.06.2023**

| Activity | Time | User's thoughts | Success/Fail |
|---|---|---|---|
| 1- Opining the application and send a message to a specific account | 0.45m | | ✅ |
| 2- As a mentor, find the mentee you are matched with. | 0.46m | | ✅ |
| 3- Find "Manage your relationship" page and hand book. | ~1m | | ❌ |
| 4- Book a meeting with your mentee. | 1.22m | Booking a meeting was really confusing | ✅ |
| 5- Add a new skill to the list of your skills | ~1m | | ✅ |
| 6- Find "Who is Re:Coded" artical. | 2.20m | I haven't even see where is this | ❌ |
| 7- Join the meeting you have created earlier. | 2.20m | | ✅ |

## User Testing

User ID: **User #6**
Test date: **27.06.2023**

| Activity | Time | User's thoughts | Success/Fail |
|---|---|---|---|
| 1- Opining the application and send a message to a specific account | 0.30m | | ✅ |
| 2- As a mentor, find the mentee you are matched with. | 1.22m | It is hard for me to understand the relationship between the mentor and the mentee | ❌ |
| 3- Find "Manage your relationship" page and hand book. | 1.24m | | ❌ |
| 4- Book a meeting with your mentee. | 2m | | ❌ |
| 5- Add a new skill to the list of your skills | 0.16m | | ✅ |
| 6- Find "Who is Re:Coded" artical. | 0.20m | | ✅ |
| 7- Join the meeting you have created earlier. | ~1m | | ❌ |

## User Testing

User ID
**User #3**

Test date
**26.06.2023**

| Activity | Time | User's thoughts | Success/Fail |
|---|---|---|---|
| 1- Opining the application and send a message to a specific account | 0.35m | | ✅ |
| 2- As a mentor, find the mentee you are matched with. | 1.3m | I couldn't understand what kind of actions I can have with my mentee. | ✅ |
| 3- Find "Manage your relationship" page and hand book. | ~1m | | ❌ |
| 4- Book a meeting with your mentee. | 1.40m | | ✅ |
| 5- Add a new skill to the list of your skills | 0.33m | | ✅ |
| 6- Find "Who is Re:Coded" artical. | 0.42m | | ✅ |
| 7- Join the meeting you have created earlier. | ~1m | | ✅ |

## User Testing

User ID
**User #4**

Test date
**27.06.2023**

| Activity | Time | User's thoughts | Success/Fail |
|---|---|---|---|
| 1- Opining the application and send a message to a specific account | ~1m | | ✅ |
| 2- As a mentor, find the mentee you are matched with. | 0.15m | | ✅ |
| 3- Find "Manage your relationship" page and hand book. | 2.25m | | ✅ |
| 4- Book a meeting with your mentee. | 1.17m | | ✅ |
| 5- Add a new skill to the list of your skills | 0.55m | | ✅ |
| 6- Find "Who is Re:Coded" artical. | 0.20m | | ✅ |
| 7- Join the meeting you have created earlier. | ~1.50m | | ✅ |

## User Testing

User ID
**User #1**

Test date
**26.06.2023**

| Activity | Time | User's thoughts | Success/Fail |
|---|---|---|---|
| 1- Opining the application and send a message to a specific account | ~2.30m | Struggling with finding the way sending a message | ✅ |
| 2- As a mentor, find the mentee you are matched with. | 1.3m | It is confusing to find the mentee. | ✅ |
| 3- Find "Manage your relationship" page and hand book. | ~1m | The Dashboard is so confusing. | ✅ |
| 4- Book a meeting with your mentee. | 0.50m | | ✅ |
| 5- Add a new skill to the list of your skills | 0.40m | | ✅ |
| 6- Find "Who is Re:Coded" artical. | ~1m | The Icons of the app are confusing | ✅ |
| 7- Join the meeting you have created earlier. | ~1m | The Dashboard needs to be more organized, I thought I can find the meeting there. | ✅ |

## User Testing

User ID
**User #2**

Test date
**26.06.2023**

| Activity | Time | User's thoughts | Success/Fail |
|---|---|---|---|
| 1- Opining the application and send a message to a specific account | 1.25m | It was quite easy to open the app. | ✅ |
| 2- As a mentor, find the mentee you are matched with. | 3m | I was so hard finding the mentee | ✅ |
| 3- Find "Manage your relationship" page and hand book. | 0.50m | | ✅ |
| 4- Book a meeting with your mentee. | 3.30m | The calendar looks confusing, I was expecting something like Google Calendar. | ✅ |
| 5- Add a new skill to the list of your skills | 0.37m | | ✅ |
| 6- Find "Who is Re:Coded" artical. | 0.20m | | ✅ |
| 7- Join the meeting you have created earlier. | ~1m | | ✅ |